\documentclass[pdflatex,sn-mathphys-num]{sn-jnl}
\usepackage{mathrsfs}
\usepackage{graphicx}%
\usepackage{multirow}%
\usepackage{amsmath,amssymb,amsfonts}%
\usepackage{amsthm}%
\usepackage[title]{appendix}%
\usepackage{lmodern}
\usepackage{textcomp}%
\usepackage{manyfoot}%
\usepackage{booktabs}%
\usepackage{caption}
\usepackage[normalem]{ulem}
\usepackage{algorithm}%
\usepackage{algorithmicx}%
\usepackage{algpseudocode}%
\usepackage{listings}%
\usepackage{graphicx}

\usepackage{setspace}

\usepackage{xurl}

\usepackage{caption}

\newcounter{extfig}

\begin{document}

\title{\vspace{-2cm}{A geometry aware framework enhances noninvasive mapping of whole human brain dynamics}}

\author[1]{\fnm{Song} \sur{Wang}}\equalcont{These authors contributed equally to this work.}

\author[1,2]{\fnm{Kexin} \sur{Lou}}\equalcont{These authors contributed equally to this work.}

\author[1,3]{\fnm{Chen} \sur{Wei}}\equalcont{These authors contributed equally to this work.}

\author[4]{\fnm{Zhiyuan} \sur{Sheng}}

\author[1]{\fnm{Jiahao} \sur{Tang}}

\author[1]{\fnm{Kaining} \sur{Peng}}

\author[1]{\fnm{Xinke} \sur{Shen}}

\author[4]{\fnm{Shuhao} \sur{Mei}}

\author*[4]{\fnm{Liang} \sur{Chen}}\email{hschenliang@fudan.edu.cn}

\author*[5]{\fnm{Dongfeng} \sur{Gu}}\email{gudf@sustech.edu.cn}

\author*[1]{\fnm{Quanying} \sur{Liu}}\email{liuqy@sustech.edu.cn}

\affil*[1]{\orgdiv{Department of Biomedical Engineering}, \orgname{Southern University of Science and Technology}, \orgaddress{\city{Shenzhen}, \postcode{518055}, \country{China}}}

\affil[2]{\orgdiv{School of Electrical Engineering and Computer Science}, \orgname{The University of Queensland}, \orgaddress{\city{Brisbane}, \postcode{4072},  \country{Australia}}}

\affil[3]{\orgdiv{School of Psychology}, \orgname{University of Birmingham}, \orgaddress{\street{Edgbaston}, \city{Birmingham}, \postcode{B15 2TT}, \country{United Kingdom}}}

\affil[4]{\orgdiv{Department of Neurosurgery, Neurosurgical Institute of Fudan University}, \orgname{Huashan Hospital}, \orgaddress{\city{Shanghai}, \postcode{200040}, \country{China}}}

\affil[5]{\orgdiv{School of Public Health and Emergency Management}, \orgname{Southern University of Science and Technology}, \orgaddress{\city{Shenzhen}, \postcode{518055}, \country{China}}}

\abstract{{Non invasive electrophysiology lacks methods that accurately reconstruct whole brain spatiotemporal dynamics while incorporating individual cortical geometry, leaving current electroencephalography and magnetoencephalography source imaging limited by simplistic or biologically implausible priors.  Here, we show that embedding patient-specific Geometric Basis Function (GBF),  eigenmodes derived from each individual’s cortical surface, provides a powerful anatomic constraint that resolves the inverse problem and improves reconstruction fidelity. The method allows reconstruction of the sources as linear combinations of geometric organization of neural dynamics. We validate GBF across the Meta-Source Benchmark, task-evoked data, resting-state networks, intracranial stimulation, and epilepsy data. The results demonstrate that GBF yields high localization accuracy and captures fast spatiotemporal dynamics consistent with anatomical pathways. These findings suggest that both spontaneous and evoked whole-brain activity can be described by hundreds of geometric modes, providing a compact yet accurate representation of neural sources. By linking cortical geometry to electrophysiological dynamics, GBF offers a versatile source imaging tool for both scientific and clinical applications.}}

\maketitle

\onehalfspacing 


\section{Introduction}\label{sec1} 

Brain waves are a core feature of neural activity, reflecting synchronized oscillations across different brain regions~\cite{baillet2017magnetoencephalography,vidaurre2018spontaneous,adamantidis2019oscillating}. These waves underpin a wide range of functions, from sensory processing and motor control to complex cognitive tasks~\cite{shine2016dynamics,vogels2005neural}. By organizing brain activity into large-scale networks, coordinated brain waves support essential processes such as attention~\cite{gaillard2022neural}, memory consolidation~\cite{staresina2024coupled}, reasoning~\cite{xu2023interacting}, and consciousness~\cite{raut2021global}. Understanding brain wave dynamics is thus fundamental to neuroscience. Given the high temporal dynamics of brain waves~\cite{timofeev2020spatio,liang2021cortex}, capturing them requires whole-brain imaging with both high spatial and temporal resolution.
This capability is essential for understanding how the brain coordinates complex functions across networks~\cite{edelman2015eeg,weiner2023propofol,coito2016directed} and for identifying abnormal dynamics in conditions like epilepsy~\cite{coito2016directed,sun2022deep,cao2022virtual}. Such imaging advances could transform our ability to map large-scale resting-state networks, pinpoint task-related areas, and delineate pathological networks. Therefore, there is a pressing need for a technique that can reveal whole-brain waves with both high temporal and spatial accuracy. Achieving this will {substantially} enhance our understanding of brain dynamics and improve clinical diagnostics and targeted treatments.

Unraveling fast-changing brain waves requires an ideal functional imaging technique with two key characteristics: high temporal resolution and whole-brain coverage. These features are critical for accurately tracking the rapid propagation of brain waves across large-scale networks.
While functional magnetic resonance imaging (fMRI) offers excellent spatial resolution and whole-brain coverage, it is an indirect measure of neural activity and lacks the temporal resolution needed to capture fast electrophysiological signals~\cite{he2008multimodal,katwal2013measuring,hall2014relationship}. Meanwhile, invasive electrophysiological recordings, such as electrocorticography (ECoG)~\cite{buzsaki2012origin} and stereoelectroencephalography (SEEG)~\cite{jaber2024spatial}, offer high temporal precision but can only capture localized areas, leaving much of the brain unobserved. This gap creates a major challenge in imaging whole-brain electrophysiological dynamics with both comprehensive coverage and high spatiotemporal resolution. Non-invasive techniques like electroencephalography (EEG) and magnetoencephalography (MEG) offer both high temporal resolution and whole-brain coverage, positioning them as promising tools for studying the large-scale communications of brain waves. 
However, the spatial resolution of EEG/MEG in sensor space is inevitably limited by the number and density of available sensors~\cite{wens2023exploring}. This underlines the importance of EEG/MEG source imaging (ESI), which aims to improve spatial resolution by using computational modeling to reconstruct whole-brain sources~\cite{liu2018detecting,sohrabpour2020noninvasive,sun2022deep}.
As multiple source configurations can explain the same sensor signals, ESI is an ill-posed problem~\cite{he2018electrophysiological}. Addressing this issue requires incorporating prior information about neural sources~\cite{zhang2025artificial}. Yet many methods rely on simplistic or non-biological priors~\cite{haufe2011large,petrov2012harmony,chowdhury2013meg}, limiting the accuracy and interpretability of the source reconstructions. Thus, advances in biologically informed priors and source modeling are crucial for capturing the complexity of brain waves and improving the reconstruction of whole-brain dynamics.

Brain waves are shaped by the brain's structural organization, including the geometry of the cortical surface and connectivity patterns~\cite{pang2023geometric,zilles2013development,koller2024human,markello2022neuromaps}. The propagation of neural activity is constrained by cortical folding, which defines the brain's surface geometry~\cite{zilles2013development}, and by complementary structural and organizational features such as white matter connections and whole-brain functional gradients~\cite{koller2024human,markello2022neuromaps}. Recent studies have shown that whole-brain dynamics observed with fMRI can be effectively described by geometric basis functions (GBFs)—the Laplace–Beltrami eigenmodes computed on each individual's cortical surface obtained from structural MRI~\cite{pang2023geometric,gabay2017cortical,cabral2023intrinsic,atasoy2018harmonic}. Remarkably, it has been reported that the whole-brain activity, observed in 32,000 fMRI vertices, can be efficiently represented by merely 200 eigenmodes through a linear combination~\cite{pang2023geometric}. This finding highlights the potential for using a small number of participant-specific geometric modes to represent large-scale brain dynamics. Given that EEG and MEG signals predominantly reflect postsynaptic cortical currents~\cite{gross2019magnetoencephalography,thio2023relative}, incorporating these same geometric bases as spatial priors for ESI offers a principled approach to enhance reconstruction accuracy and neurophysiological validity.

Here, we introduce the GBF framework, a geometry-constrained approach for EEG/MEG source imaging that reconstructs neural activity in alignment with each brain's intrinsic cortical geometry (Fig.~\ref{fig:fig1}).
Unlike conventional ESI methods that rely on simplistic priors, GBF directly integrates individualized cortical structure derived from MRI, providing anatomically grounded constraints for source imaging.
We validate GBF through a series of analyses, including synthetic benchmarks, task-based EEG, functional connectivity (FC) reconstruction, intracranial stimulation, and clinical epilepsy localization.
These analyses collectively demonstrate that embedding geometric structure into source imaging enables biologically interpretable and spatially precise reconstruction of large-scale neural dynamics, linking cortical anatomy with functional activity in both cognitive and clinical contexts.

\section{Results}\label{sec2}
\subsection{Geometric basis functions constrained EEG source reconstruction}

\begin{figure}[p]
    \centering
    \captionsetup{
    font=footnotesize,
    labelfont=bf,
    textfont=normalfont,
    skip=3pt}
    \includegraphics[width=\textwidth]{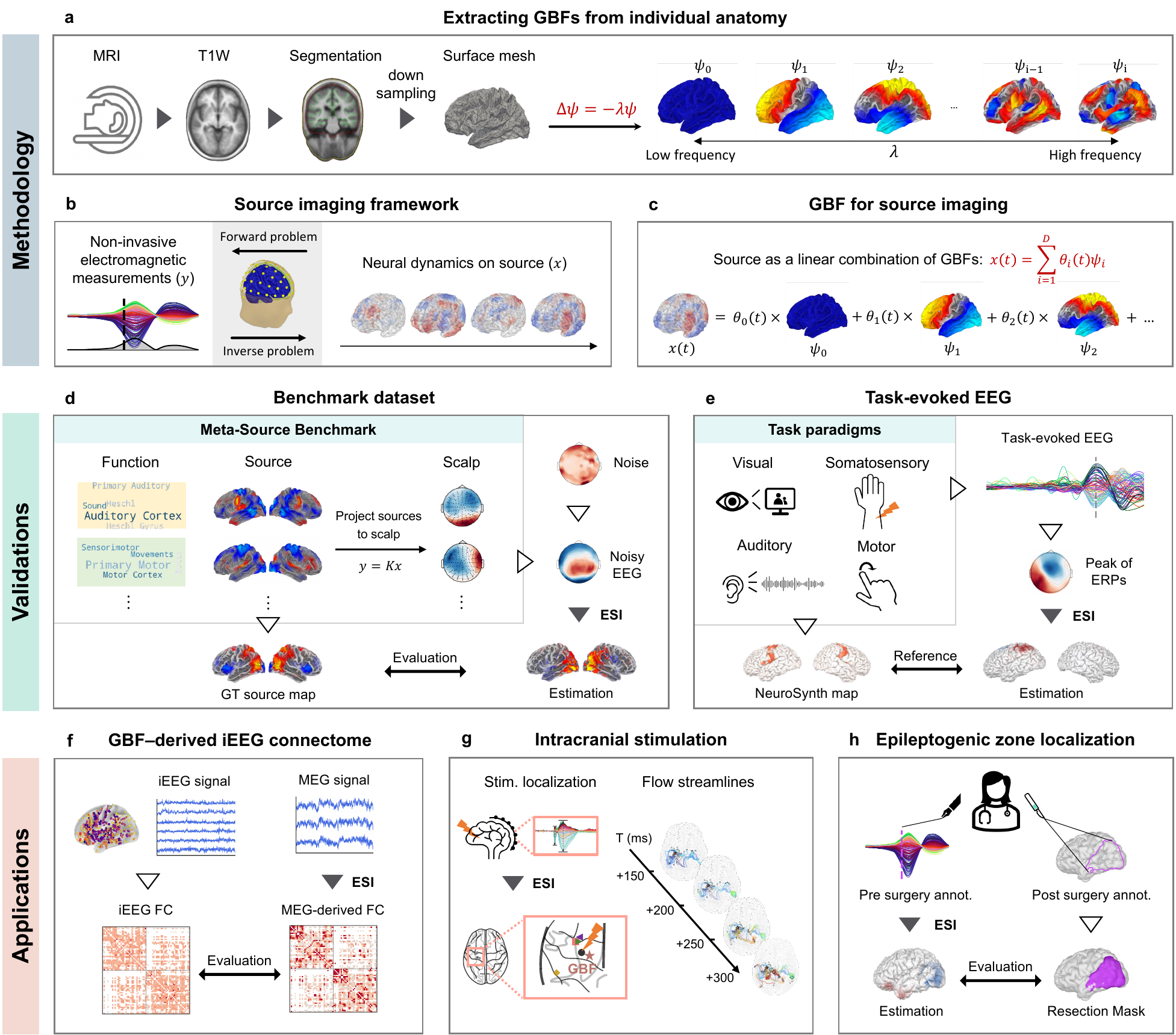}
    \caption{\textbf{Geometric basis function framework for EEG/MEG source imaging.}
    \textbf{a.} Geometric basis functions (GBFs) are derived from individual cortical surface meshes obtained from structural MRI by solving the Laplace–Beltrami eigenvalue problem, $\Delta\psi = -\lambda\psi$. The eigenmodes are ordered from low to high spatial frequency, representing long- to short-wavelength geometric patterns of the cortex. Values are colour-coded from negative to positive, ranging from light blue to yellow. 
    \textbf{b.} In the ESI framework, scalp potentials $y$ and cortical sources $x$ are linked through the forward model $y = Kx$, where $K$ is the lead-field matrix describing the head's volume conduction. 
    \textbf{c.} In our Geometric basis function (GBF) framework, the inverse problem is solved under geometric constraints by expressing cortical sources as linear combinations of GBFs, $x(t)=\sum_i\theta_i(t)\psi_i$, thereby embedding cortical geometry directly into the inverse operator for biologically informed source reconstruction. 
    \textbf{d.} The Meta-Source Benchmark was generated from large-scale meta-analytic fMRI maps (NeuroVault–NeuroSynth), projected onto the cortical surface to create realistic source–EEG pairs across sensory and motor domains. EEG signals were simulated by adding noise to scalp projections, and reconstructed sources were compared with the ground truth for evaluation.
    \textbf{e.} Task-evoked EEG data from visual, auditory, somatosensory and motor paradigms were used to extract event-related potentials, which were reconstructed into cortical source maps. The resulting activations were compared with task-specific meta-analytic Neurosynth maps for spatial validation.
    \textbf{f.} Group-level intracranial EEG (iEEG) recordings were used to construct a whole-brain functional connectome. Resting-state MEG signals were source reconstructed to the corresponding cortical locations, and the resulting functional connectivity (FC) matrices were compared with the iEEG connectome to assess cross-modal correspondence.
    \textbf{g.} Scalp EEG recorded during intracranial electrical stimulation (iES) was analyzed with GBF to reconstruct cortical responses to direct perturbations. The framework localizes stimulation sites on individual cortical surfaces and tracks the spatiotemporal propagation of evoked activity using phase-gradient optical flow analysis.
    \textbf{h.} Interictal spikes from clinical EEG were identified and marked by clinicians before surgery. GBF reconstructed cortical sources to infer epileptogenic zones (EZ) from interictal spikes, which were then compared with postoperative resection masks defined by surgeons.
    }
    \label{fig:fig1}
\end{figure}

GBF integrated individual cortical geometry into source reconstruction, embedding biologically informed constraints that align neural dynamics with the brain's intrinsic structure (Fig.~\ref{fig:fig1}a–c). 
Specifically, the cortical surface is reconstructed from individual MRI scans, then decomposed into Laplace–Beltrami eigenmodes, resulting in geometric basis functions that span a range of spatial scales from smooth to fine-grained (Fig.~\ref{fig:fig1}a).

In the GBF framework, cortical sources $x(t)$ are represented as linear combinations of geometric basis functions $\psi_i$, i.e., $x(t)=\sum_i \theta_i(t)\psi_i$, providing a geometry-informed prior that constrains the inverse problem.

We validated the performance and generalizability of GBF by assessing its ability for source localization (Fig.~\ref{fig:fig1}d, e). We first used synthetic EEG data with known ground truth sources for quantitative evaluation (Fig.~\ref{fig:fig1}d), and then tested GBF's capacity to recover canonical sensory and motor activations from task-evoked EEG data (Fig.~\ref{fig:fig1}e), further demonstrating its spatial precision.

Finally, we highlighted the potential of brain source-level analyses beyond GBF-based source localization (Fig.~\ref{fig:fig1}f–h). 
For instance, GBF was applied to investigate large-scale functional connectivity in MEG source space, revealing resting-state brain networks with strong alignment to intracranial EEG (iEEG) results (Fig.~\ref{fig:fig1}f). Additionally, we used GBF to study the effects of intracranial electrical stimulation (iES) on neural activity and its propagation dynamics (Fig.~\ref{fig:fig1}g), and explored the clinical potential of GBF by applying it to epilepsy localization (Fig.~\ref{fig:fig1}h). 
These analyses not only validate GBF's accuracy in source reconstruction but also showcase its potential to advance our understanding of brain function and its clinical applications.

\subsection{Source localization on Meta-Source Benchmark}

\begin{figure}[ht]
    \centering
    \includegraphics[width=1\textwidth]{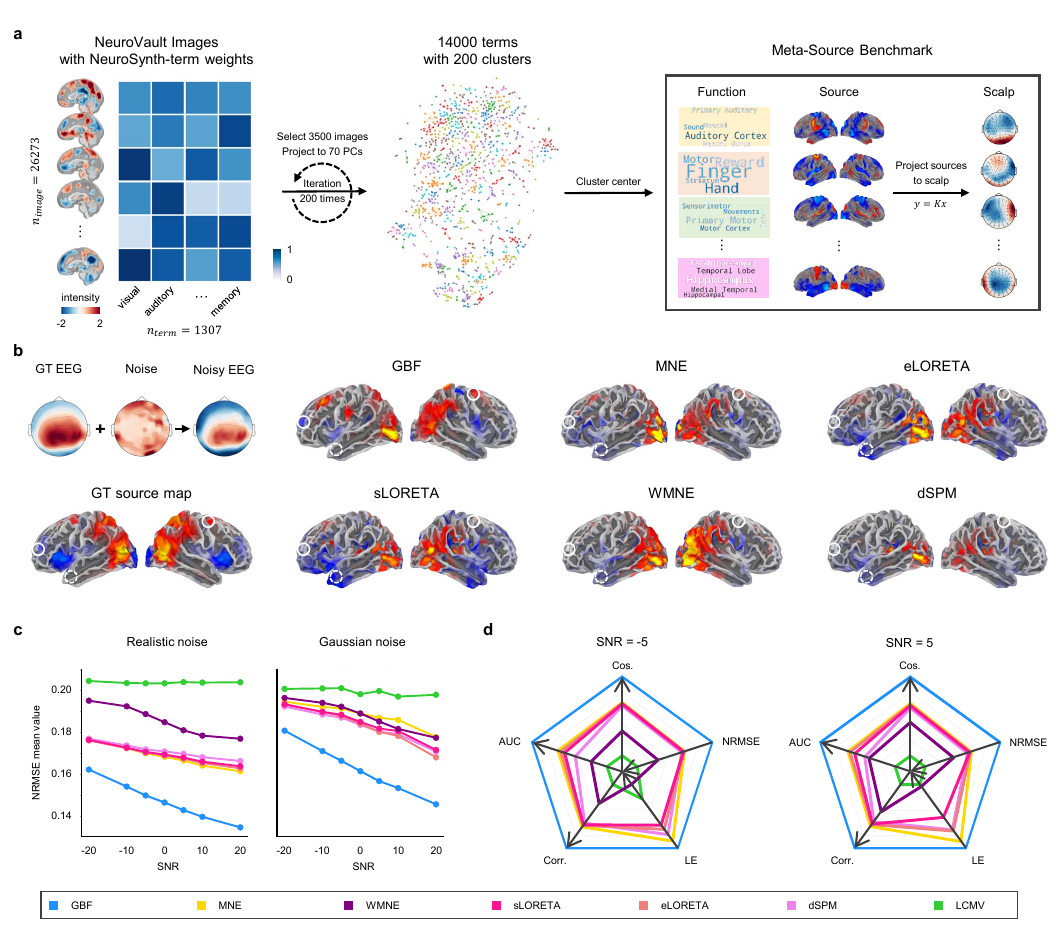}
    \caption{\textbf{Generating Meta-source Benchmark dataset and comparing ESI methods under various noise conditions.} 
    \textbf{a.} The EEG source-scalp paired data were generated by fMRI meta-analysis from NeuroVault images, clustered based on word vectors, then transformed from volume space (MNI) to surface space (fsaverage), and finally projected onto the EEG scalp. 
    \textbf{b.} Ground truth spatial maps and corresponding scalp signals for an example component with realistic noise are shown. ESI results from GBF are compared against other ESI methods (displayed in color). 
    \textbf{c.} Method performance under different noise conditions for various ESI methods, evaluated by NRMSE. 
    \textbf{d.} Performance comparison of different ESI methods at SNR = 5, assessed using five evaluation metrics: Cosine Similarity (Cos.), AUC, Pearson Correlation (Corr.), Localization Error (LE), and NRMSE.}
    \label{fig:sim_framework}
\end{figure}

Reliable evaluation of source localization methods requires well-defined benchmark datasets. In real EEG data, the true spatiotemporal distribution of neural sources is typically unknown, making direct validation challenging~\cite{qin2023evaluation,reynaud2024comprehensive,wang2024advancing}. To address this, synthetic source-EEG datasets are commonly generated using head models with predefined source ~\cite{allouch2023effect,leone2024investigating}.
Traditional synthetic datasets often rely on either randomly positioned isolated activation seeds~\cite{Long-Short-Term} or localized patches representing regional activations~\cite{Sparse-Bayesian-Learning,Data-Synthesis-Based}. These approaches fail to capture the complexity of real brain activity patterns and lack biological interpretability. To overcome these limitations, we developed a framework to systematically generate source–EEG pairs and their functional descriptors (Fig.~\ref{fig:sim_framework}a; details in Method~\ref{method:Meta-source benchmark}). 
This benchmark consists of 200 high-resolution source maps paired with their corresponding scalp EEG topographies. 
By linking statistical fMRI meta-analytic maps with Neurosynth cognitive descriptors, it bridges cognitive functions and their spatial representations in the brain~\cite{beam2021data,pacella2024morphospace}. 
These meta-analytic maps serve as biologically informed spatial references for evaluating source localization methods under realistic distributed configurations.

We applied GBF to reconstruct neural sources from benchmark EEG data and assessed accuracy using five complementary metrics: NRMSE, localization error, Pearson correlation, cosine similarity, and AUC (Supplementary Note 1.3 for details). To assess the robustness of GBF under different noise conditions, we introduced Gaussian noise and realistic noise extracted from real EEG recordings (Supplementary Note 1.1). Fig.~\ref{fig:sim_framework}b illustrates the estimated source maps at SNR = 5 with realistic noise, demonstrating that GBF closely matched the ground truth. {Regional error and depth-related analyses are provided in Extended Fig.~1. Representative examples of benchmark source classes and their corresponding reconstructions across methods are shown in Extended Fig.~2.}
We further compared GBF with conventional source localization methods, including Minimum Norm Estimation (MNE), weighted MNE (wMNE), sLORETA, and dSPM. The results indicate that GBF outperforms conventional methods in accuracy and robustness. {Further comparisons are provided in Extended Fig.~3 and Supplementary Figs~5-6, 11.}
Fig.~\ref{fig:sim_framework}c demonstrates the performance of ESI methods under realistic noise and Gaussian noise from the perspective of the mean value of NRMSE for different SNR. Fig.~\ref{fig:sim_framework}d presents a detailed visualization of their performance across five metrics with realistic noise (Detailed results in Supplementary Table 6 and 7). Statistical significance was assessed using paired t-tests or Wilcoxon tests based on normality, with FDR correction applied. Across all evaluation metrics, the GBF significantly outperformed other approaches (Supplementary Table 10 and 11). 

Although GBF achieved the best overall accuracy, it failed to reproduce the ground-truth spatial pattern in all regions (Fig.~2B). Region-wise normalized error map at SNR = 5 dB (real noise) indicates that residuals primarily occur in deep/ventral regions, aligning with depth-related sensitivity and volume-conduction constraints (Extended Fig.1 a-b). Nonetheless, GBF yielded the lowest global NRMSE and lower regional NRMSE in approximately 95\% of parcels (Extended Fig.1 c-d). Two factors likely drive the remaining discrepancies: (i) the ill-posed nature of the inverse problem and measurement noise, whereby many GBF coefficient combinations project similarly to the sensors, and (ii) basis truncation, as the first 200 modes capture about 87\% of cortical variance (Supplementary Fig.~9), while including higher-order modes improves noiseless fits but worsens conditioning and degrades performance under noise (See Extended Fig.~1f).

\subsection{GBF captures task-evoked neural patterns}

\begin{figure}[ht!]
    \centering
    \includegraphics[width=1\textwidth]{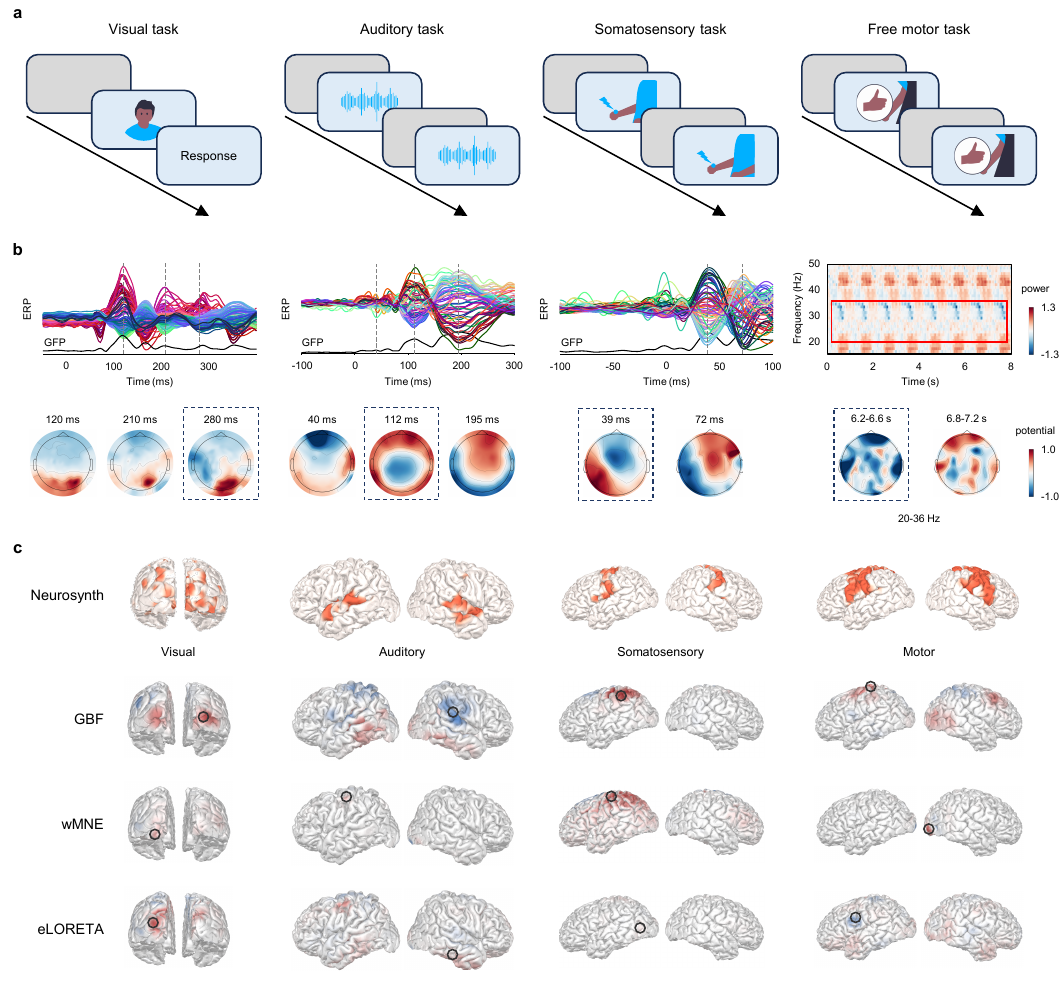}
    \caption{\textbf{Task-based EEG source analysis.}
    \textbf{a.} Task paradigms for each study: Visual task with a 300 ms stimulus presentation followed by a response window; Auditory task with tones of approximately 100 ms presented every 1.7 seconds; Somatosensory task with 0.2 ms median nerve stimuli presented every 1.7 seconds; and a self-paced free motor task with a 3-second inter-trial interval.
    \textbf{b.} ERPs and GFP responses for visual, auditory, and somatosensory tasks, showing time series of ERPs with corresponding GFP peaks. ERD/ERS for the free motor task is shown as baseline-normalized power (log-ratio; baseline period, $-4$ to $-3.5$ s). Scalp topographies at selected time points illustrate characteristic activity patterns across tasks. Colours in the ERP plots denote different EEG electrodes; colour scales are given in log-ratio for the time–frequency map and in arbitrary units (a.u.) for the scalp topographies.
    \textbf{c.} Source localization results using various methods, compared to task-related meta-analytic maps from Neurosynth which are transformed into each participant's native surface space for anatomical alignment. Source estimates are presented for GBF, WMNE, and eLORETA. Each row displays estimated cortical activation maps. Black circles indicate peak activation locations identified by each method.
    }
    \label{fig:task_results}
\end{figure}

To examine the generalizability of GBF to real experimental settings, we analyzed EEG recordings from well-established cognitive tasks (see Section~\ref{method:Task-related EEG analysis}).
Specifically, we reconstructed EEG sources during visual, auditory, somatosensory, and motor tasks (Fig.~\ref{fig:task_results}a).  For each task, stimulus-locked responses were extracted to analyze temporal dynamics and scalp topographies  (Fig.~\ref{fig:task_results}b), demonstrating that scalp topographies at selected latencies revealed task-specific neural activation patterns that are well aligned with established neurophysiological findings. 
Fig.~\ref{fig:task_results}c compared source localization results using GBF, wMNE, and eLORETA. Activation maps reconstructed with GBF closely matched the reference maps from the Neurosynth meta‑analysis, which were transformed into each {participant}'s native surface space to ensure anatomical correspondence. This alignment enabled a more direct comparison across methods. Notably, during the visual task, GBF accurately localized activity within the primary visual cortex, reflecting expected anatomical distribution. Similarly, for the auditory task, GBF effectively pinpointed bilateral auditory cortices, consistent with known auditory processing regions. During somatosensory stimulation (i.e., median nerve stimulation to the right hand), GBF localized the peak activation in the contralateral somatosensory cortex, well matched with established somatotopic organization. Likewise, in the motor task (i.e., tapping the right middle finger), GBF reliably captured periodic modulation of cortical activity within the left motor cortex. These findings collectively underscore the improved spatial localization accuracy achieved by GBF relative to conventional methods. We assessed the source level time courses reconstructed by GBF for each task and each {participant}, GBF consistently produced clear stimulus-evoked peaks followed by a subsequent decay, and the resulting temporal profiles were closely aligned with the stimulus onset (Supplementary Figs. 14–19). In contrast, conventional methods demonstrated greater variability and less precise temporal dynamics. These results underscore that the GBF method outperforms conventional approaches in capturing task-evoked neural activity, providing substantial improvements in both spatial localization accuracy and temporal resolution.

\subsection{GBF–derived MEG virtual iEEG connectomes recapitulate intracranial resting‐state organization}
\begin{figure}[!htbp]
    \centering
    \captionsetup{
    font=footnotesize,
    labelfont=bf,
    textfont=normalfont,
    skip=3pt}
     \includegraphics[width=0.92\textwidth]{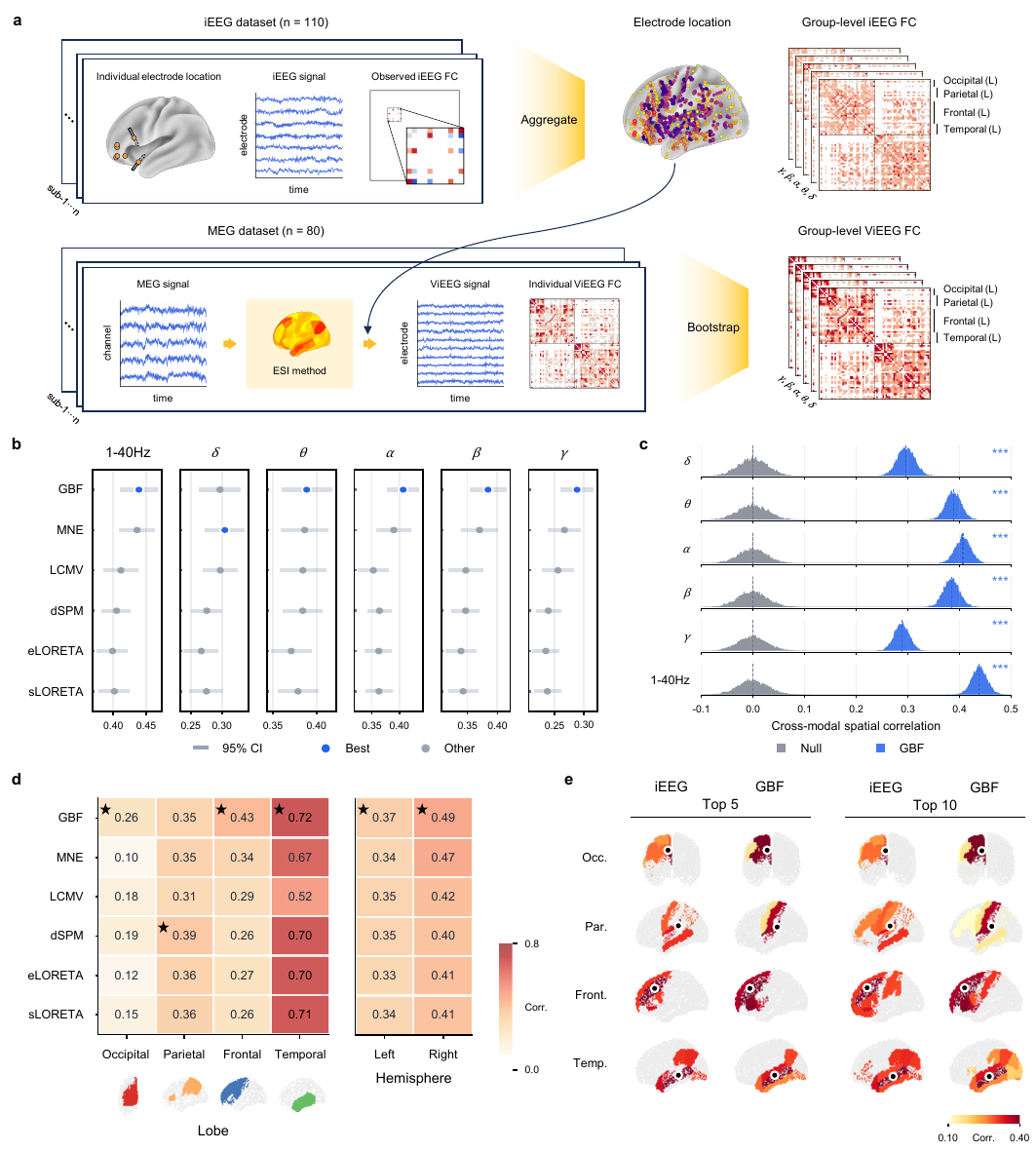}
    \caption{\textbf{Validation of MEG-derived ViEEG connectomes against group iEEG connectivity}
    \textbf{a.} Pipeline. Upper panel: In the iEEG dataset (n = 110), bipolar contacts are assigned to atlas ROIs. Patient-level ROI-to-ROI functional connectivity is computed and then averaged across patients to generate a group iEEG connectome. Lower panel: For each HCP participant (n = 80), MEG data are source-reconstructed with the candidate ESI method, projected to the same atlas contacts to yield ViEEG time series, and individual FC matrices are built. Group ViEEG connectomes are estimated using 5,000 bootstrap resamples that match the per-edge sampling of the iEEG dataset.
    \textbf{b.} {Pearson correlations between the group iEEG connectome from 110 patients and group ViEEG functional connectivity derived from MEG data of 80 participants across the broadband range (1--40 Hz) and five canonical frequency bands. Group-level correlations were estimated from 5,000 bootstrap resamples matched to the per-edge sampling of the iEEG dataset. Points, mean correlations across bootstrap resamples; whiskers, 95\% bootstrap confidence intervals; blue points, best-performing inverse method in each frequency band.}
    \textbf{c.} {Comparison of GBF correlation distributions against a sampling-matched null model. For each frequency band, the bootstrap distribution of GBF correlations (blue; 5,000 resamples) was compared with a sampling-matched null distribution (grey; 5,000 resamples). Blue asterisks (***) denote adjusted P $<$ 0.001 in a two-sided overlap-based test after Benjamini--Hochberg correction (all adjusted \(P = 4.0 \times 10^{-4}\)). Dashed lines represent medians.}
    \textbf{d.} Mean iEEG–ViEEG correlations within each lobe (Occipital, Parietal, Frontal, Temporal) and hemisphere in $\beta$ band. Higher values indicate better recovery of region-specific connectivity; the star marks the leading score in each panel.
    \textbf{e.} Edge-level alignment ($\beta$ band). For a representative seed ROI (Left-Cuneus; Left-Parietal operculum; Left-Middle frontal gyrus; Left-Middle temporal gyrus) in each lobe (black circle), the top-5 and top-10 strongest iEEG edges (left column) are shown alongside the corresponding edges derived from GBF-ViEEG (right column). Corr., Pearson correlation.}
    \label{fig:fig4_ieeg}
\end{figure}

Resting-state functional connectivity captures large-scale communication in the brain and reflects its intrinsic network architecture~\cite{arnulfo2020long,bassett2017network,margulies2016situating}. Group-level FC matrices ($\geqslant$ 40 {participants}) exhibit stable, reproducible organization~\cite{ma2024effect} and show convergent features across neuroimaging modalities~\cite{Betzel2019,afnan2025validating}. To test whether geometry-constrained inversion more faithfully recovers this architecture from MEG, we compared MEG-derived FC against two references: a within-modality electrophysiological benchmark from iEEG (See Fig.~\ref{fig:fig4_ieeg}) and a cross-modal benchmark from fMRI BOLD (Extended Data Fig.~4). Methods for FC construction are detailed in Methods~\ref{method:Vieeg_ieeg} and Supplementary Note 3-4.

Resting-state iEEG FC was derived from pre-surgical {patients with epilepsy} (n = 110) ~\cite{Frauscher2018}. MEG FC was computed from the HCP dataset (n = 80) (Fig.~\ref{fig:fig4_ieeg}a). Across the broadband range (1--40 Hz) and the $\delta$, $\theta$, $\alpha$, $\beta$, and low-$\gamma$ bands, spatial correlations between the group ViEEG and iEEG connectomes were moderate (Pearson's \(r = 0.20\text{--}0.45\); Fig.~\ref{fig:fig4_ieeg}b). GBF matched MNE in the $\delta$ band and outperformed all alternative methods in the $\theta$, $\alpha$, $\beta$, and low-$\gamma$ bands. Relative to a sampling-matched spatial null {(randomized participant--edge assignments preserving per-edge sampling counts)}, GBF correlation distributions were consistently shifted above null in all bands {(two-sided overlap-based test; all FDR-adjusted \(P = 4.0 \times 10^{-4}\); Fig.~\ref{fig:fig4_ieeg}c)}, indicating genuine cross-dataset agreement rather than sampling or spatial-autocorrelation artifacts. Performance peaked in $\beta$, where GBF achieved the highest lobe/hemisphere-resolved correlations (occipital, parietal, frontal, temporal) and reproduced the spatial layout of high-strength iEEG edges (top-$k$ co-localized edges per lobe; Fig.~\ref{fig:fig4_ieeg}d–e).
In contrast, cross-modal correlations between functional connectivity derived from MEG and fMRI showed systematically lower across all methods (Extended data Fig.~4, Supplementary Fig.~13), as expected given differences introduced by neurovascular coupling and spatial blurring. Under these cross-modal conditions, GBF performed comparably to other methods.
In sum, in the within-modality electrophysiology comparison, GBF more faithfully recovers intrinsic electrophysiological connectivity. By contrast, cross-modal correlations between MEG- and fMRI-derived connectivity are modest across all approaches, whereas GBF leads in most electrophysiology-based comparisons across frequency bands and cortical territories.

\subsection{GBF identifies stimulation sites and delineates the spatial-temporal dynamics in intracranial stimulation}

We further investigated the applicability of the GBF in capturing evoked cortical dynamics under controlled perturbations. Specifically, we applied GBF to a dataset of simultaneous iES and high-density scalp EEG recordings~\cite{mikulan2020simultaneous}. An effective source imaging approach should enable both the accurate localization of stimulation sites and the reconstruction of the ensuing spatiotemporal patterns of neural activity, thereby capturing the full trajectory of stimulation-induced cortical responses~\cite{unnwongse2023validating,alberto2021meg}.

\begin{figure}[!htbp]
    \centering
    \captionsetup{
    font=footnotesize,
    labelfont=bf,
    textfont=normalfont,
    skip=3pt}
    \includegraphics[width=0.95\textwidth]{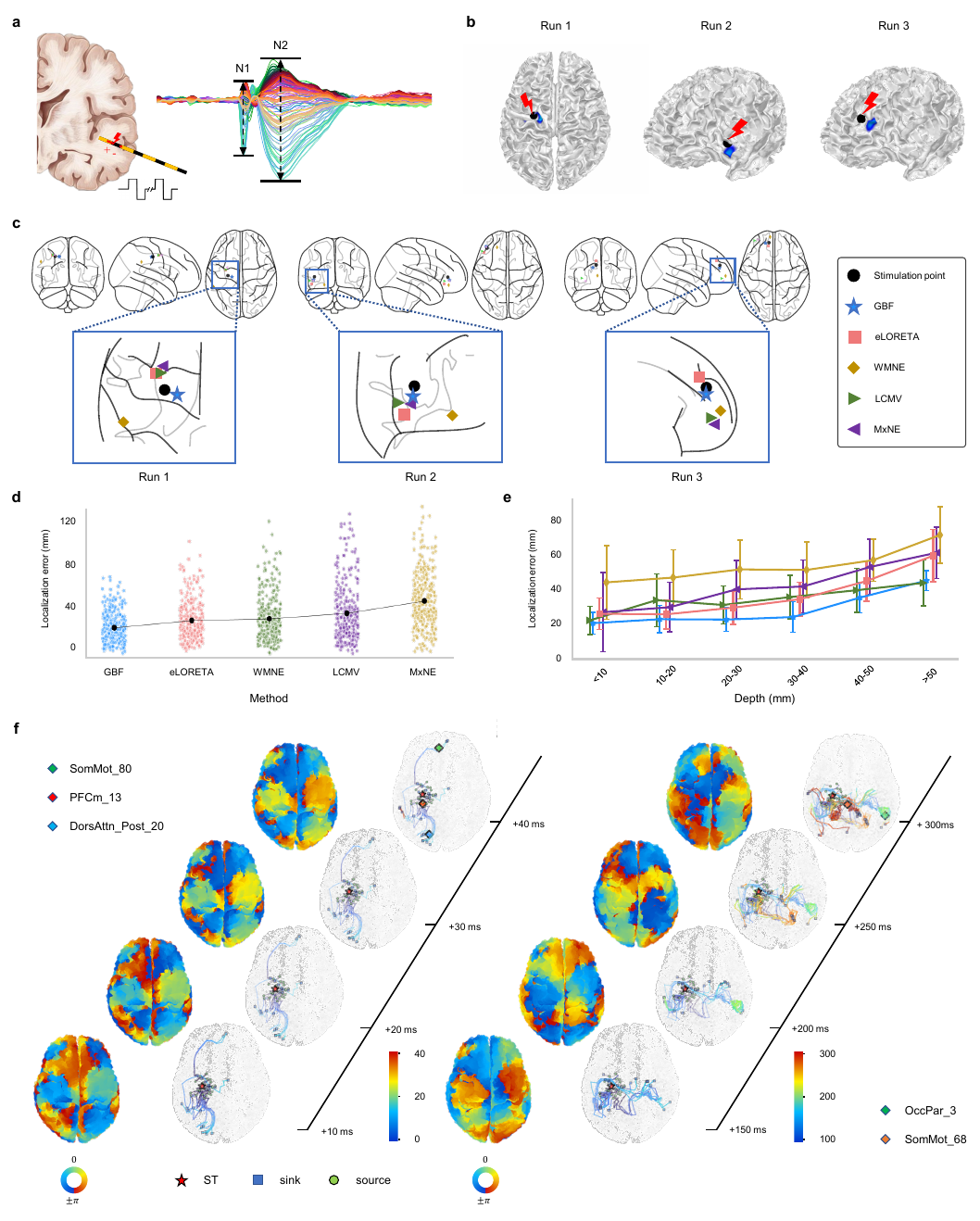}
    \caption{\textbf{Spatiotemporal neural responses to intracranial stimulation.}  
    \textbf{a.} Schematic of the intracranial stimulation setup, illustrating the stimulation point and the corresponding evoked electrophysiological response, highlighting the N1 and N2 components.  
    \textbf{b.} Example of source localization results for {participant 07}, Runs 1, 2, 3 using GBF. Thresholded stimulation-evoked activity is shown in blue; black dots mark the ground-truth intracranial stimulation sites. 
    \textbf{c.} Localization results across three runs using different methods. The true stimulation point is marked in black, while each method's estimated location is color-coded according to the legend. The insets provide zoomed-in views of the estimated locations within the cortical space.  
    \textbf{d.} {Localization error for each inverse method across all stimulation sessions (n = 318). Each colored dot represents a session, and black dots indicate the mean localization error for each method.}
    \textbf{e.} {Depth-binned comparison of localization error across inverse methods. Sessions with valid depth assignments were grouped into 10-mm bins: \textless 10 mm, 10--20 mm, 20--30 mm, 30--40 mm, 40--50 mm, and 50--60 mm, with \(n =\) 9, 99, 84, 56, 48 and 13, respectively (total \(n = 309\) sessions). Points, mean localization error within each depth bin; error bars, s.d.}
    \textbf{f.} GBF-based reconstruction of phase maps and optical-flow streamlines following intracranial stimulation. Streamlines trace the instantaneous flow field; clustered endpoints mark convergence hubs. The red star marks the stimulation target; green and blue denote source and sink, respectively; diamonds indicate parcels crossed by the streamlines. Panels are ordered by latency (N1 early, N2 late). Circular phase wheel (0–2$\pi$) encodes instantaneous phase; linear colorbar encodes streamline travel time (ms).}
    \label{fig:ccep_results}
\end{figure}

Fig.~\ref{fig:ccep_results}a demonstrates the experimental setup, in which intracranial electrodes deliver electrical stimulation pulses that generate evoked signals (N1 and N2) in scalp EEG. The cortical stimulation site was determined following the procedure described in Method \ref{method:CCEP}, and reconstruction accuracy was quantified by measuring the spatial distance between the ground truth and estimated stimulation sites.
Fig.~\ref{fig:ccep_results}b illustrates representative examples of GBF for one {participant} across three sessions, showing thresholded stimulation-induced activity and the ground-truth intracranial stimulation locations.
Fig.~\ref{fig:ccep_results}c visualizes the reconstructed peak sites obtained from GBF and four alternative methods (wMNE, eLORETA, sLORETA, and LCMV) for the corresponding sessions in a glass brain view. Ground-truth stimulation points are represented as black dots, and reconstructed points are indicated using colored symbols. The enlarged views illustrate that the GBF method provides consistently superior reconstructions, closely aligned with the actual intracranial stimulation sites. To systematically quantify localization accuracy, we calculated localization errors in all sessions and {participants}, and compared GBF against the four alternative methods (Fig.~\ref{fig:ccep_results}d).
All pairwise method comparisons were statistically significant (Wilcoxon signed-rank test, FDR corrected; p\textsubscript{max} = 2.42$\times$10\textsuperscript{--9}).
To examine robustness across stimulation depths, we stratified sessions into depth bins (Fig.~\ref{fig:ccep_results}e; Supplementary Note 5.2; Supplementary Tables 13--14). {A linear mixed-effects model showed that localization error increased with stimulation depth (\(\beta = 0.428\), \(P = 2.83 \times 10^{-5}\)) and remained lower for GBF than for eLORETA, LCMV, MxNE and wMNE at the mean stimulation depth (\(\beta = 6.926, 8.624, 13.832,\) and \(25.671\); \(P = 1.86 \times 10^{-5}, 9.82 \times 10^{-8}, 1.24 \times 10^{-17},\) and \(1.09 \times 10^{-56}\), respectively). The depth-related increase in error was steeper for eLORETA and MxNE than for GBF (\(\beta = 0.296\) and \(0.394\); \(P = 0.0300\) and \(0.00394\)), but did not differ significantly for LCMV or wMNE (\(\beta = -0.142\) and \(-0.022\); \(p = 0.2979\) and \(0.8745\)).}

In addition to stimulation location, we investigated the spatiotemporal propagation of cortical responses induced by intracranial stimulation. A representative session ({participant} 07, Run 05) was selected as a case study to illustrate GBF's capability in capturing neural dynamics (Fig.~\ref{fig:ccep_results}f). We computed phase gradients across cortical regions within the N1 and N2 windows and derived optical-flow vectors and streamlines to track propagation (See Method~\ref{method:wave_optical_flow} for details)~\cite{roberts2019metastable}. During the N1 interval, GBF captured rapid, millisecond-scale, directed propagation initiating from the stimulation site, and converging on three clusters (SomMot\_80, PFCm\_13 and DorsAttn\_Post\_20). During the N2 interval, propagation was broader and slower, with sustained spread toward clusters (OccPar\_3 and SomMot\_68). {We further quantified wave propagation in the N1 and N2 windows using direction consistency, propagation speed, and streamline path length (Supplementary Table~15).} Similar patterns were observed in two additional {participants} (Supplementary Figs. 22-23). These findings align closely with previous CCEP studies describing sequential cortical responses following intracranial stimulation~\cite{veit2021temporal, lemarechal2022brain}.

\subsection{GBF facilitates epileptogenic zone localization in epilepsy}

The epileptogenic zone (EZ) is the minimal cortical area indispensable for seizure generation; seizure free typically requires its complete removal or disconnection~\cite{rosenow2001presurgical,zijlmans2019changing}. Localizing the EZ from EEG is a key step in presurgical planning, informing candidacy, resection extent, and expected outcomes~\cite{sebastiano2020identifying}. In contrast, the seizure-onset zone (SOZ) is the region where ictal activity originates and is commonly used as the best clinical estimate of EZ~\cite{li2018using}. Nonetheless, the SOZ and EZ are distinct entities, as removing the SOZ does not always result in seizure freedom~\cite{tamilia2017current}. We evaluated the use of GBF to localize EZ and SOZ in clinical settings, measuring their alignment with surgical resections as the gold standard for EZ and comparing with clinician-identified SOZs.

\begin{figure}[!ht]
    \centering
    \includegraphics[width=1\textwidth]{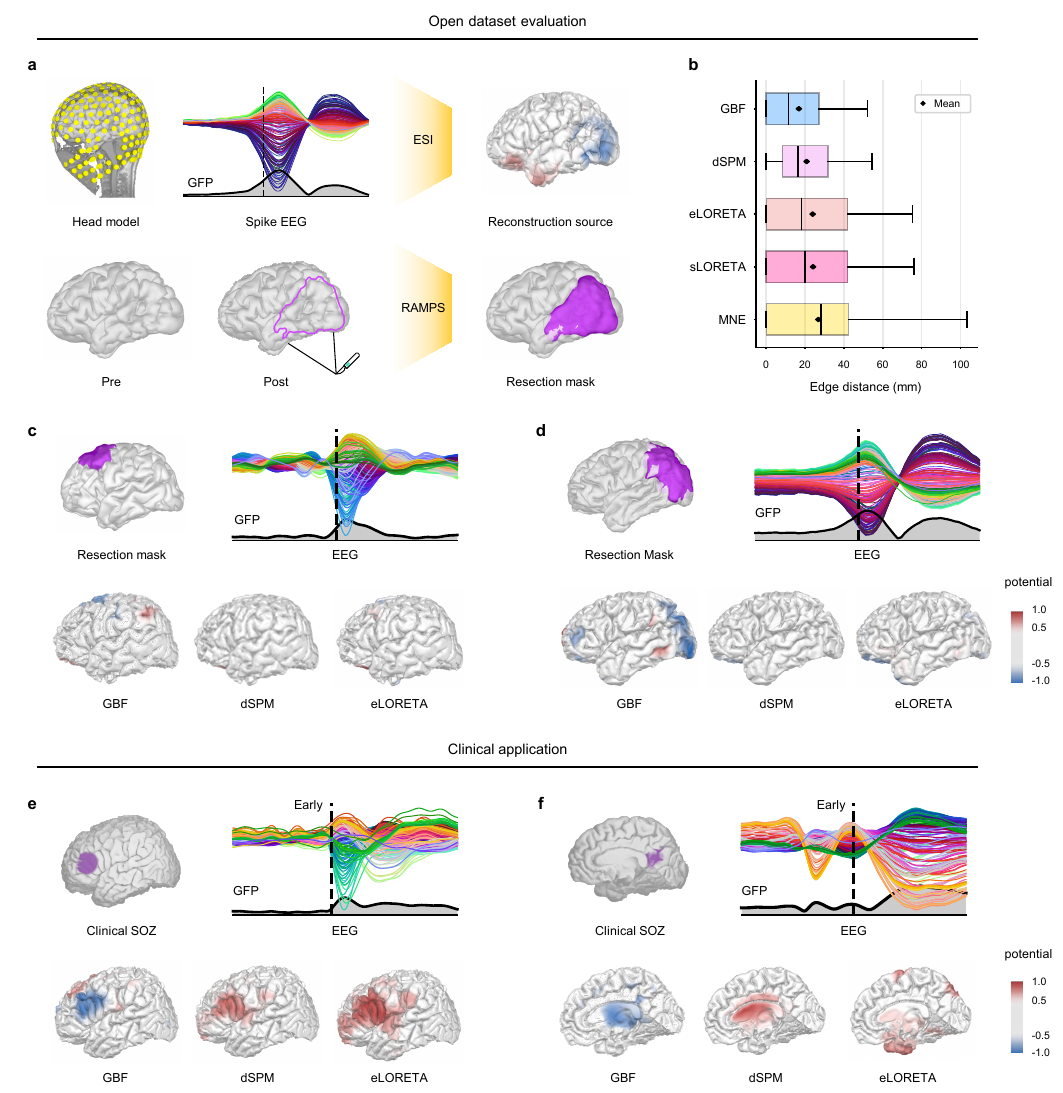}
    \caption
    { 
    \textbf{GBF for epileptogenic zone localization in epilepsy.}
    \textbf{a.} {Patient-specific} head models is reconstructed from T1-weighted MRI. Interictal spikes are time-locked using the global field power (GFP). Cortical sources are reconstructed at the mid-GFP frame (dashed line; the first time point at which GFP reaches 50 $\%$ of its peak). {Patient-specific} resection cavities are reconstructed from paired pre-/post-operative T1 volumes with RAMPS, are overlaid for visual comparison (purple).
    \textbf{b.} {Minimum Euclidean distance from the vertex of maximal activation to the resection-mask boundary (0 mm if the maximum lies inside; n=24, ILAE 1–2).} Boxes show median and IQR; whiskers, 1.5×IQR; diamonds, means. 
    \textbf{c-d.} Two representative patients (Patient 02 and Patient 09) with resection masks and sensor-level GFP (mid-GFP frame marked). The cortical maps at this frame are reconstructed using GBF, dSPM, and eLORETA.
    \textbf{e-f.} Two additional {patients} from Huashan Hospital with clinician-defined SOZs (purple) and reconstructions at the early time point. Color bars indicate relative activation on a common scale.
    }
    \label{fig:epi_results}
\end{figure}

We first conducted a resectional ground-truth validation on a presurgical dataset with EEG and MRI (description and preprocessing in Supplementary Note 8.5)~\cite{vorderwulbecke2025high}. We evaluated patients with favorable 12-month postsurgical outcomes (n $=$ 24), analyzed interictal spikes, and reconstructed patient-specific resection cavities using RAMPS~\cite{simpson2025automated}. Fig.~\ref{fig:epi_results}a outlines the cohort and processing pipeline (details in Methods~\ref{method:EZ}; Supplementary Note~6.1). Localization accuracy was defined as the minimum Euclidean distance from the vertex of maximal activation to the boundary of the patient-specific resection mask (assigned \(0\,\mathrm{mm}\) if the peak lay inside the mask). Across patients, GBF yielded smaller distances and a significant paired two-sided improvement over MNE (p = 0.012; Fig.~\ref{fig:epi_results}b). Relative to other alternative methods, GBF showed consistent but non-significant advantages (p = 0.079, 0.128, 0.133 for dSPM, sLORETA, eLORETA, respectively). Fig.~\ref{fig:epi_results}c-d show representative {patients} and viewpoints; with additional {patients} and multi-view visualizations in Supplementary Fig.~33.

To evaluate seizure onset localization rather than the typically broader resection cavities, we further analyzed two patients with 256-channel HD-EEG from Huashan Hospital and two patients with publicly available 19-channel low-density EEG~\cite{zwolinski2010open} (Supplementary Fig.~25). In both cases, from Huashan and public datasets, the GBF peak was closer to the clinician-identified SOZ compared to other estimations, with early-frame maps displaying focal patterns consistent with annotated onsets (Fig.~\ref{fig:epi_results}e-f). Frequency-domain Granger causality was also computed among GBF-identified early nodes (See Supplementary Figs.~25-30, Note 6), uncovering causal links in neural oscillations.
Together, these findings establish GBF as a geometry-aware and clinically applicable method for noninvasive EZ and SOZ identification, enhancing alignment with surgical ground truth.

\section{Discussion}

EEG/MEG source imaging has long been limited by the ill-posed nature of the inverse problem, leading to blurred spatial maps and complicating biological interpretation~\cite{da2013eeg,he2018electrophysiological}. In this study, we addressed this challenge by introducing Geometric Basis Functions, which directly embed each individual's cortical geometry into the EEG/MEG inverse solution. This anatomically grounded framework enables precise reconstruction of large-scale neural dynamics, enhancing both spatial accuracy and interpretability. Through systematic experiments, including the Meta-Source Benchmark, task-based EEG validations and clinical applications, we demonstrate that the GBF approach achieves high reconstruction accuracy and recovers biologically meaningful neural patterns. It reliably localizes task-evoked sources, preserves large-scale connectivity patterns, localizes intracranial stimulation sites, tracks millisecond-scale propagation, and delineates epileptogenic zone that validated against surgical ground truth and clinical annotations. Together, these results establish the GBF framework as a robust, generalizable method for linking brain geometry to the propagation of neural waves, substantially advancing the accuracy and interpretability of noninvasive brain imaging.

A key feature of our approach is the integration of each brain's geometry as an intrinsic prior for source estimation. Participant-specific basis functions are derived from the cortical surface using Laplace–Beltrami decomposition, yielding spatial bases that capture both global patterns (low-frequency eigenmodes) and fine-grained anatomical details (high-frequency eigenmodes)~\cite{pang2023geometric}. These geometry-informed bases guide source estimates toward spatially coherent and anatomically plausible distributions. In contrast, conventional methods often rely on simplifying assumptions or require extensive parameter tuning. For example, some approaches assume a fixed spherical manifold~\cite{petrov2012harmony}, employ hand-crafted patch priors that depend on substantial hyperparameter optimization~\cite{mattout2005multivariate}, or define connectome eigenmodes at the regional level without capturing individual cortical geometry~\cite{atasoy2016human,yeh2021mapping,xia2024decomposing,yang2023eigenmode}. Our approach improves these limitations by leveraging each individual's MRI-derived cortical surface to construct an efficient, anatomically grounded representation defined directly on the {participant}'s native surface mesh (i.e., vertex-wise mesh), rather than coarse regional parcellations. We implement a closed-form MAP solution, which is both mathematically elegant and computationally efficient. This solution incorporates a logarithmic spectral prior ($\Sigma^{-1}{=}\mathrm{diag}[-\beta/\log(\lambda_i)]$), prioritizing smooth, low-frequency modes while only moderately penalizing higher-frequency modes for numerical stability. The closed-form nature of GBF approach simplifies the computation, making it both straightforward to implement and efficient in practice. This prior structure strikes an optimal balance between biological realism and model flexibility, enabling precise reconstructions without requiring complex optimization procedures. Comparative analyses showed that ablating the geometry prior (i.e., using identity covariance) led to unstable, noisy reconstructions, while overly steep decay schemes (e.g., power law or exponential) caused over-smoothing (as detailed in Supplementary Tables 8 to 9). {Additionally, sensitivity analyses indicated that reconstruction performance plateaued at approximately 300 GBFs, whereas inclusion of substantially higher-order modes primarily increased ill-conditioning without improving effective spatial resolution (Extended Data Fig.~1f; Supplementary Fig.~9), supporting the use of this range as an empirically motivated operating point.} This finding is in line with both the theoretical spatial resolution limits of EEG/MEG~\cite{michel2012towards,baillet2017magnetoencephalography} and previous reports indicating that a small number of geometric modes (about 200) account for the majority of variance in fMRI signals~\cite{atasoy2016human,pang2023geometric}.

To rigorously assess the benefits of incorporating cortical geometry, we developed the Meta-Source Benchmark, which provides a functionally meaningful and methodologically independent testbed for ESI. Unlike traditional benchmarks relying on oversimplified point sources or Gaussian patches~\cite{wei2021edge,Sparse-Bayesian-Learning,Data-Synthesis-Based}, our Meta-Source Benchmark leverages meta-analytic fMRI activation maps linked to cognitive terms~\cite{beam2021data,pacella2024morphospace}. These maps, retrieved from NeuroVault and annotated with cognitive labels, are semantically clustered to generate biologically realistic and distributed source activation patterns (Section~\ref{method:Meta-source benchmark}; Supplementary Note~1). Although these maps are not direct electrophysiological recordings, they provide a functionally grounded and independent reference with whole‑cortex coverage and support evaluation of ESI methods. The benchmark accommodates extensive variations, including different noise patterns (e.g., Gaussian and task-derived), source space definitions (e.g., FEM and BEM), and head-model configurations (e.g., changing tissue conductivities) and sensor layouts (e.g., 32-channel EEG and 128-channel EEG), enabling comprehensive and flexible evaluation of ESI methods. Initially constructed in the fsaverage surface space, the benchmark maps can be transformed into individual participant spaces, facilitating cross‑participant validation. The current release includes 200 components (examples in Extended Data Fig.~2), and it is designed for future expansion with additional datasets and clustering strategies. {As the Meta-Source Benchmark is derived from meta-analytic fMRI spatial maps rather than electrophysiological source dynamics, it should be interpreted as a spatially informed reference for evaluating distributed inverse solutions rather than as an electrophysiological ground truth.} Within this framework, GBF demonstrated superior accuracy in reconstructing distributed benchmark reference patterns, highlighting the importance of individualized cortical geometry in ESI (Supplementary Figs.~5--7; Supplementary Tables~5--6,~10--11).

After validating the localization accuracy of the GBF method, we further explored whether geometry‑constrained inverse solutions could capture not only focal activity but also large‑scale neural dynamics, which are critical for neuroscience and clinical applications~\cite{baillet2017magnetoencephalography,vidaurre2018spontaneous,kucyi2020electrophysiological}. Specifically, in the resting-state analysis, we assessed whether the reconstructed EEG/MEG sources reflected coherent functional networks. The results showed that the spatial alignment between the EEG/MEG and fMRI networks was weaker compared to MEG-derived connectomes and iEEG. This discrepancy likely arises from neurovascular and preprocessing confounds, as well as the frequency sensitivity inherent in cross-modal interactions between EEG/MEG and fMRI~\cite{xavier2025consistency,meyer2013electrophysiological}. Next, in the iES data, GBF localized peak activity with higher precision than traditional methods. Additionally, using optical flow analysis~\cite{roberts2019metastable}, we identified stimulation-evoked traveling waves with GBF, consistent with findings from intracranial recordings~\cite{campbell2025macroscale}. {This interpretation is further supported by recent work suggesting that geometrically constrained coupling can help organize large-scale propagation patterns, supporting the idea that eigenmode-like cortical geometry may shape distributed wave dynamics~\cite{phogat2025unified}.} Further analysis revealed propagation hubs and trajectories aligned with previously documented pathways; for example, stimulating the premotor cortex propagated signals to the frontal and occipital cortices, consistent with known motor-execution pathways~\cite{zanon2013long,zhao2025progressively,busan2012transcranial}. In epilepsy datasets, applying GBF to interictal EEG recordings {enabled delineation of} the epileptogenic zone, showing strong agreement with invasively confirmed seizure-onset zones and clinical annotations. Although a small systematic offset of several centimeters was noted between the non-invasive localization of interictal spikes and the surgically determined SOZ, it is consistent with prior studies indicating that interictal spike imaging may not capture the full ictal‑onset network~\cite{abou2022noninvasive,zijlmans2019changing}. To move beyond focal localization and characterize the epileptogenic network, we applied frequency-domain Granger causality on GBF-reconstructed sources (Supplementary Note~5). This revealed directed interactions, with propagation from the SOZ to downstream regions and network hubs, consistent with previous reports~\cite{park2018granger,wang2024causal}. Together, these findings demonstrate that embedding individual cortical geometry improves source localization and enables network-level analyses of resting-state organization, stimulation-evoked propagation, and epileptogenic networks.

Several aspects of this study remain open for further exploration. 
First, our analyses primarily focus on cortical surface reconstruction, although GBF is also compatible with subcortical areas. We have demonstrated the preliminary feasibility of extending GBFs to subcortical geometry (Extended Data Fig.~5; Supplementary Note 7). However, further validation is required to assess the full capacity of GBFs in reconstructing deeper sources, as structures such as the hippocampus and thalamus play {important} roles in cognition and pathology~\cite{schucht2013subcortical,pereira2025subcortical,norden2002role,badawy2013subcortical}. 
Second, the current framework constrains sources spatially while treating each time point independently. It can be easily extended to incorporate temporal priors or dynamical models, such as state-space formulations~\cite{sohrabpour2020noninvasive,yang2016state}, which would enhance reconstruction fidelity and enable real-time tracking of neural dynamics.
Third, geometric basis functions are well-suited for integration with deep learning networks, serving as compact, participant-specific geometric representations that networks can directly use. By acting as geometric priors, GBFs can reduce parameter and data requirements while ensuring model predictions are constrained to biologically plausible manifolds~\cite{dong2025brain}.
Finally, broader validation across larger and more diverse cohorts is essential for establishing the generalizability and clinical applications of GBF. Although our results in epilepsy are promising, expanding evaluations to other neurological conditions and healthy populations will further assess the robustness of GBF. For instance, structural anomalies, such as lesions or cortical malformations~\cite{joutsa2022brain,klingler2021mapping}, may challenge the geometric assumptions underlying GBFs. To address these challenges, adaptive strategies, such as excluding lesioned regions from the mesh and recomputing GBFs from altered anatomy, could extend the framework's applicability to abnormal brain structures.
Addressing these directions will {substantially} enhance the GBF's applicability, strengthening its potential as a general tool for linking brain structure to source dynamics.

\section{Methods}

\subsection{GBF framework}
\subsubsection{Geometric basis decomposition}

An adapted Laplace–Beltrami operator is employed to characterize the intrinsic geometry of the cortical manifold embedded in three-dimensional Euclidean space \cite{seo2011laplace}. The differential operator is formally defined as:

\begin{equation}
\Delta : = \frac{1}{W}\sum_{i,j} \frac{\partial}{\partial x_i} \left( g^{ij} W \frac{\partial}{\partial x_j} \right)
\label{eq:LBO_operator }
\end{equation}

where  \(x_i, x_j\) denote local parametric coordinates on the manifold, \(g^{ij}\) represents the contravariant components of the inverse metric tensor, and \(W = \sqrt{\det(G)}\) corresponds to the volume element with \(G\)  being the matrix representation of the metric tensor \(g\). The first fundamental form  \(g\) encodes the intrinsic geometry of the cortical surface through its characterization of infinitesimal arc lengths and angular relationships. The contravariant tensor \(g^{ij}\) facilitates coordinate-invariant computations through tensor transformation laws, thereby preserving the intrinsic curvature variations and anisotropic scaling properties of the cortical manifold.

This operator generates spectral embeddings through the solution of the eigenvalue problem:

\begin{equation}
\Delta \psi_k = -\lambda_k \psi_k, \quad k=1,...,S
\label{eq:eigenproblem }
\end{equation}

where  \(\psi \in\mathbb{R}^{S \times N}\) denotes the orthonormal eigenmodes (with S = 300 modes in our implementation) and  \(\lambda \in\mathbb{R}^{S}\) represents the associated eigenvalues sorted in ascending order of spatial frequency. The eigenmodes $\psi_k$ exhibit hierarchical spatial organization, where lower eigenvalues correspond to slowly varying global patterns, while higher eigenvalues capture localized high-frequency variations. For neurobiophysical interpretation, the eigenvalue spectrum is converted into prior variance components governing the spatial weighting of cortical source estimates.

\subsubsection{Source estimation}

The source activity is modeled by the equation:

\begin{equation}
    y = K A \theta + \varepsilon,
\end{equation}

where \( A \) is the spatial basis function matrix, \( \theta \) are the source coefficients, \( K \) is the leadfield matrix, and \( \varepsilon \) represents Gaussian noise. Let \( L = K A \), where \( L \) denotes the effective lead-field matrix in the basis-function space, then the model can be rewritten as:

\begin{equation}
    y = L \theta + \varepsilon.
\end{equation}

Assuming that the source coefficients \( \theta \) follow a Gaussian distribution \( \theta \sim N(0, \Sigma) \), where \( \Sigma \) is a diagonal covariance matrix, the regularization term is introduced as:

\begin{equation}
    \Sigma^{-1} = \operatorname{diag}\!\left(-\frac{\beta}{\log \lambda}\right),
\end{equation}

which helps prevent singularities in the covariance matrix and stabilizes the inversion process. 
Here, \( \lambda \) controls spectral smoothness by penalizing high-frequency geometric modes, whereas \( \beta \) adjusts region-specific prior strength across anatomical domains. In practice, a single global \( \beta \) was used for all analyses except subcortical localization, where region-specific weighting was applied.
By applying Bayes' theorem, the posterior distribution of the source coefficients \( \theta \) is given by:

\begin{equation}
    P(\theta \mid y) \propto \exp\left( -\frac{1}{2}(y - L\theta)^T(y - L\theta) - \frac{1}{2} \theta^T \Sigma^{-1} \theta \right).
\end{equation}

The Maximum A Posteriori (MAP) estimation of \( \theta \) is obtained by maximizing the posterior distribution, which is equivalent to minimizing the negative log of the posterior:

\begin{equation}
    \theta_{\text{MAP}} = \arg\min_{\theta} \left[ (y - L\theta)^T (y - L\theta) + \theta^T \Sigma^{-1} \theta \right],
\end{equation}

and the corresponding closed-form solution is

\begin{equation}
    \hat{\theta} = \left( L^T L + \Sigma^{-1} \right)^{-1} L^T y.
\end{equation}

A concise description for other methods, including assumptions, advantages, limitations and implementation details could be found in Supplementary Table 1

\subsection{Meta-source benchmark}\label{method:Meta-source benchmark}

We developed a comprehensive data generation framework that utilizes spatial maps derived from fMRI meta-analyses to create a realistic synthetic dataset for validating ESI algorithms (Fig. \ref{fig:sim_framework}a). Specifically, we randomly selected 26,273 statistical images from NeuroVault~\cite{gorgolewski2015neurovault}, which were transformed into the MNI-152 volume space and annotated with 1,307 Neurosynth terms~\cite{yarkoni2011large}. Each term represents a specific functional theme, encapsulating brain regions associated with particular cognitive processes or behaviors. The value associated with each term represents the degree to which a statistical image reflects activation related to a given cognitive function or experimental condition as identified in Neurosynth's large-scale text-mining analysis of neuroimaging studies. These functional terms are derived from a large-scale database of cognitive neuroscience studies, creating a semantic link between brain activity and cognitive function. For example, a term like "visual processing" may encompass brain regions activated during tasks involving visual stimuli.

From this annotated set, we constructed an image-term matrix that reflects the relationship between brain activity patterns (as represented by the statistical images) and their corresponding cognitive themes (as represented by the Neurosynth terms). Then we applied Principal Component Analysis (PCA) to reduce the dimensionality and extract the most relevant features.

In each iteration of the PCA, 70 spatial components were extracted from the image-term matrix using the Nilearn library. This process was repeated in 200 iterations, yielding a total of 14,000 spatial components. These components were then clustered into 200 unique spatial maps using a K-Nearest Neighbors (KNN) clustering algorithm, based on their shared topics, determined by the associated functional terms. This clustering process reflects shared cognitive and functional themes across brain activity patterns, providing a representation of large-scale brain processes commonly observed in cognitive tasks. 

The clustered spatial maps were subsequently transformed into the fsaverage surface space using Neuromaps~\cite{markello2022neuromaps, wu2018accurate}. These final spatial maps, now in surface space, served as the source spatial maps in the benchmark. To generate EEG sensor data, we performed forward modeling using participant-specific head models. Each surface-based source map was projected to EEG scalp topographies via the boundary element method (BEM) forward model. To simulate realistic conditions, we added two types of noise: (1) Gaussian white noise at controlled signal-to-noise ratios (SNR), and (2) realistic noise sampled from empirical EEG covariance matrices (see Supplementary Note 1 for details). The resulting dataset comprises 200 source-EEG pairs with known ground-truth source locations and functional descriptors.

\subsection{Datasets from human participants}\label{method:datasets}
This study utilized multiple publicly available and clinically acquired datasets to demonstrate the performance of GBF-based source localization across a variety of tasks and clinical conditions. The datasets encompass cognitive tasks, resting-state activity, iES, and SOZ localization (further details in Supplementary Note 8). Additionally, a private clinical dataset from Huashan Hospital, with labels provided by medical experts, was included.
\textbf{Cognitive Tasks}: The VEPCON dataset~\cite{pascucci2022source} included 128-channel high-density EEG and structural MRI data from 20 healthy participants performing face/motion visual discrimination tasks.
The OSE dataset~\cite{yamashita2024opm} provided 63-channel EEG from 5 healthy participants across auditory, voluntary movement, and median-nerve stimulation tasks.
\textbf{Resting-State Connectivity}: The HCP dataset~\cite{van2013wu} provided MEG, structural MRI, and resting-state fMRI data from 100 healthy participants.
The LEMON dataset~\cite{babayan2019mind} included 62-channel EEG data from 110 participants.
Intracranial EEG data were sourced from the MNI Open iEEG Atlas~\cite{frauscher2018atlas}, which contains recordings from 110 {patients with epilepsy} (1,715 bipolar iEEG channels, 68 seconds of eyes-closed resting-state recordings).
\textbf{Intracranial Electrical Stimulation}: The iES-CCEP EEG dataset~\cite{parmigiani2022simultaneous} featured 156-channel EEG data from 37 participants (35 retained after quality checks) across 318 sessions, involving intracranial electrical stimulation.
\textbf{SOZ Localization}: The HDEEG-IED-SurgOutcome dataset~\cite{vorderwulbecke2025high} included averaged interictal discharges from 257-channel EEG and MRI data from 24 patients with favorable surgical outcomes (ILAE 1/2). The eeg.pl dataset~\cite{zwolinski2010open} contained preoperative 19-channel EEG and MRI data from two patients with identified IED epochs.
Additionally, a private dataset from Huashan Hospital consisted of 256-channel EEG and MRI data from two patients with interictal spikes, manually annotated and validated by expert clinicians.

Individual sMRI data were processed using FreeSurfer via Neurodesk to extract both cortical and subcortical structures~\cite{fischl2012freesurfer,renton2024neurodesk}. The resulting surfaces were used to construct a three-layer (scalp, skull, brain; conductivities 0.30/0.006/0.30\,S/m) volume conductor model via the BEM in MNE-Python~\cite{gramfort2013meg}, with the sensor layout co-registered to the scalp surface. The cortical surface defined the source space with orientation-fixed dipoles aligned to the local cortical normal (approximately 10,000 sources per hemisphere for real data; and about 4,000 per hemisphere for benchmarks, down-sampled for computational efficiency). The forward (lead-field) matrix $K \in \mathbb{R}^{M \times N}$  was then computed, describing how activity at \(N\) sources projects to \(M\) sensors. Both the sensor data and $K$ were pre-whitened using the noise covariance, and the same lead field was used identically across all inverse methods. A schematic of this pipeline is provided in Supplementary Fig.~4

The cortical surfaces were further decomposed into GBFs for inverse computation.
EEG data preprocessing including bandpass filtering , bad channel interpolation, re-referencing, and independent component analysis (ICA) to remove artifacts. For task-based EEG, epochs were extracted around stimulus events, and baseline correction was applied. Resting-state EEG was segmented into 2s non-overlapping windows for source reconstruction and functional connectivity analysis (Details in Supplementary Note 8).

\subsection{Task-related EEG analysis}
\label{method:Task-related EEG analysis}

We performed source localization on task-evoked EEG responses from three stimulus-driven paradigms—visual, auditory, and somatosensory. Raw EEG was band-pass filtered and notch-filtered to remove noise, followed by inspection for bad channels and transient artifacts (details in Supplementary Note 8). Data were epoched relative to stimulus onset and baseline-corrected using the pre-stimulus interval. Trials from the same condition were then averaged to obtain condition-specific ERPs with improved signal-to-noise ratio.

Motor trials were processed separately because the task predominantly elicited event-related (de)synchronization (ERD/ERS) rather than a strictly phase-locked ERP. Motor-related data were therefore filtered in the relevant sensorimotor bands, segmented around movement onset, and analyzed in the time–frequency domain to derive ERD/ERS contrasts, which were subsequently used as input to the GBF framework (see Supplementary Note 2).

\subsection{iEEG and ViEEG connectome analysis}\label{method:Vieeg_ieeg}

Resting-state iEEG FC was derived from 110 pre-surgical {patients with epilepsy}~\cite{Frauscher2018}, aggregating bipolar channels free of interictal epileptiform discharges (1,715 channels total; 68~s eyes-closed rest per channel, resampled to 200~Hz, band-pass filtered 0.5--80~Hz, notch-filtered at 50/60~Hz). Electrodes were mapped to 76 cortical ROIs in the parcellation distributed through the Medical Image Computing and Computer Assisted Intervention (MICCAI) 2012 Grand Challenge and Workshop on Multi-Atlas Labeling (38 per hemisphere; subcortical nodes excluded). ROI--ROI edges were computed as the average AEC across inter-ROI bipolar channel pairs within patients, retaining only edges observed in $\geq$2 patients (620 edges: 271 left intra-hemispheric, 262 right intra-hemispheric, 87 inter-hemispheric).

For MEG participants (HCP dataset, n = 80), ViEEG was estimated at standardized cortical contact coordinates from the MNI Open iEEG atlas~\cite{Frauscher2018}, mapped to each {participant}'s native anatomy via affine registration (DIPY's \texttt{AffineRegistration} with mutual-information objective~\cite{garyfallidis2014dipy}). ViEEG potentials were computed from MEG source time series assuming an infinite homogeneous medium (conductivity $\sigma=0.25$~S~m$^{-1}$), with distances $<4$~mm clipped to 4~mm to avoid near-field instability. ViEEG was preprocessed by per-channel $z$-scoring and removal of the first principal component to suppress drifts and global artifacts.

To match iEEG's multi-patient edge sampling, 5,000 bootstrap MEG connectomes were constructed by randomly selecting participants to mirror per-edge patient counts in iEEG. AEC-based connectivity was estimated in $\delta$ (0.5--4~Hz), $\theta$ (4--8~Hz), $\alpha$ (8--13~Hz), $\beta$ (13--30~Hz), $\gamma$ (30--70~Hz), and broadband (1--40~Hz) over full segments (13,600 samples at 200~Hz for iEEG; 12,000 for MEG). ROI--ROI values were averaged over channel pairs bridging ROIs. MEG--iEEG agreement was quantified via Pearson correlations across the 5,000 bootstraps, tested against a spatially matched null (5,000 ROI-label permutations; significance at $<$2.5\% overlap between observed and null distributions, two-sided 5\% criterion). Details in Supplementary Note 3.

\subsection{iES data analysis}\label{method:CCEP}

\subsubsection{Electrode registration and stimulation site identification}
To evaluate the accuracy of reconstructed stimulation sites, we analyzed 318 stimulation sessions from 35 participants, using simultaneous iES and high-density scalp EEG recordings~\cite{mikulan2020simultaneous}. Electrode coordinates provided in MNI space were first mapped to individual source space following the method of Withers et al.~\cite{withers2023identifying}. For each session, the ground-truth stimulation site was defined as the cortical location closest to the actual stimulation electrode, and the reconstructed site was determined by identifying the peak source activity within a ±5 ms window centered on the recorded stimulation time.

For iES electrode registration, standard electrode positions were converted to RAS (Right-Anterior-Superior) space. KDTree structures were constructed separately for the left and right hemispheres in the source space. The stimulation site was assigned to the nearest cortical vertex from both hemispheres, and the corresponding hemisphere was recorded as the stimulation hemisphere.

To estimate stimulation depth, a KDTree was also built using the scalp vertices from the boundary element model. For each stimulation point, the shortest distance to the scalp surface along the normal direction was calculated. This distance was used to quantify depth, while avoiding confounds from non-cranial regions such as the spine or chin. Further details on electrode projection and stimulation depth estimation are provided in Supplementary Note 5.

\subsubsection{Wave propagation and optical flow analysis}\label{method:wave_optical_flow}
To quantify the propagation of stimulation-evoked waves, a local phase-velocity vector field was calculated at each time point following Roberts et al.~\cite{roberts2019metastable}. To improve computational efficiency and obtain smoother, noise-robust propagation fields, surface-level activity was aggregated to the regional (parcel) level using the Schaefer-500 atlas~\cite{schaefer2018local}. First, the instantaneous phase $\phi(x,y,z,t)$ at each node/parcel was obtained by the Hilbert transform. The local velocity vector was then defined as $- \left( \left| \frac{\partial \phi}{\partial t} \right| / \left\| \nabla \phi \right\|^{2} \right) \nabla \phi$. The spatial phase gradients on the cortical mesh were estimated using a discrete (surface-based) gradient operator~\cite{illoul2011some}, and the temporal derivatives were calculated by first-order finite differences. To avoid phase-unwrapping artifacts, we evaluated the gradient of the complex phase $e^{i\phi(x,y,z,t)}$ and used the identity $\frac{\partial \phi}{\partial x} = -i e^{-i\phi} \frac{\partial}{\partial x} e^{i\phi}$. Vector fields were interpolated between nodes with constrained natural-neighbor interpolation~\cite{illoul2011some}. Flow streamlines were traced by forward Euler integration from each node (step length 8\,\text{mm}); integration terminated when a streamline exited the brain mask or no further vector could be defined. To identify dominant pathways, we first discarded very short streamlines ($<$20 steps), then retained the last five points of each remaining streamline to capture asymptotic behaviour, and clustered these points with DBSCAN~\cite{ester1996density} (radius = 12\,\text{mm}, minimum cluster size = 10). Clusters were interpreted as dense flow convergence regions.

\subsection{Epileptogenic zone analysis}\label{method:EZ}

We analyzed patients with favorable 12-month outcomes (ILAE~1–2; \(n=24\)) after excluding cases that failed EEG or resection-mask quality control. Patient-specific resection cavities were reconstructed from paired pre- and postoperative T1-weighted MRI using RAMPS~\cite{simpson2025automated}. For the averaged interictal spike, sources were estimated at the mid-GFP frame. The mid-GFP was defined as the earliest time point at which the global field power reached \(50\%\) of its peak identified in each participant (Supplementary Note~6.1).

For SOZ localization, sources were additionally reconstructed at an early time on the GFP rising edge to target the initial generators clinically annotated as the SOZ~\cite{plummer2019interictal} (Supplementary Note~6.2). Finally, frequency-domain Granger causality was used to quantify directed interactions between the SOZ and other cortical regions (Supplementary Note~6.3).

\subsection{Ethics Statement for {patients with epilepsy} at Huashan Hospital}

EEG data from {patients with epilepsy} were collected at the Neurosurgical Institute of Fudan University, Huashan Hospital. All experimental procedures were approved by the Ethics Committee of Huashan Hospital, Fudan University (Huashan Institutional Review Board, HIRB). Written informed consent was obtained from all participants or their legal guardians prior to data acquisition, in accordance with the Declaration of Helsinki. All clinical data were anonymized before analysis to ensure patient privacy.

\section{Data availability}

The data supporting the findings of this study are available from the following public resources: the HCP dataset (\url{https://www.humanconnectome.org/study/hcp-young-adult/document/1200-subjects-data-release}), the VEPCON dataset (\url{https://openneuro.org/datasets/ds003505/versions/1.1.1}), the OSE dataset (\url{https://vbmeg.atr.jp/nictitaku209/}), the MNI Open iEEG Atlas (\url{https://mni-open-ieegatlas.research.mcgill.ca/}), the LEMON dataset (\url{https://ftp.gwdg.de/pub/misc/MPI-Leipzig_Mind-Brain-Body-LEMON}), the iES-CCEP EEG dataset (\url{https://osf.io/wsgzp/}), and the HDEEG-IED-SurgOutcome dataset (\url{https://doi.org/10.25493/B3B8-XPM}). Raw patient data from Huashan Hospital are not publicly available owing to patient privacy protections and institutional restrictions. De-identified processed neurophysiological and neuroanatomical data from the Huashan cohort are available from the corresponding authors upon reasonable request.

\section{Code availability}

The code is publicly available at \url{https://github.com/ncclab-sustech/GBFs}

\section{Acknowledgements}

This work was supported by the National Natural Science Foundation of China (62472206), National Key R\&D Program of China (2025YFC3410000), Shenzhen Science and Technology Innovation Committee (RCYX20231211090405003, RCBS20231211090748082, KJZD20230923115221044), General Program of Guangdong Natural Science Foundation (2026A1515010121), Guangdong Provincial Key Laboratory of Advanced Biomaterials (2022B1212010003), and the open research fund of the Guangdong Provincial Key Laboratory of Mathematical and Neural Dynamical Systems, the Center for Computational Science and Engineering at Southern University of Science and Technology.

\section{Author Contributions}

{Q.L. conceived and supervised the study. S.W., K.L., C.W., and Q.L. designed the study and the algorithm architecture, performed the experiments, analysed the data, and wrote the manuscript. Z.S., S.M., and L.C. provided the clinical data and contributed to data annotation and analysis. J.T. developed the code, maintained the open-source platform, and performed data preprocessing. K.P. and X.S. contributed to manuscript preparation and figure generation. D.G. contributed to clinical data acquisition and data support. All authors discussed the results and commented on the manuscript. S.W., K.L., and C.W. contributed equally to this work.}

\section{Competing Interests}

All authors declare no competing interests.

\clearpage

\bibliography{sn-bibliography}

\end{document}